\documentclass[12pt,a4paper]{article}
\pdfoutput=1
\usepackage{amssymb,amsmath,enumerate,mathrsfs,enumerate,comment}
\usepackage{array}
\usepackage{jheppub}

\usepackage{hypcap}
\usepackage[]{hyperref}
\usepackage{url}
\usepackage{filecontents}
\usepackage{placeins}
\usepackage{afterpage}

\hypersetup{
  bookmarks=true,         
  unicode=false,          
  pdftoolbar=true,        
 pdfmenubar=true,        
 pdffitwindow=true,     
 pdfstartview={FitH},    
 pdfsubject={GUT and Topological defects},   
 pdfnewwindow=true,      
 pdfcreator={RevTeX},
 colorlinks=true,       
 linkcolor=blue,          
 citecolor=blue,        
 filecolor=black,      
 urlcolor=blue,           
  }

\allowdisplaybreaks

\renewcommand\afterAffiliationSpace{\vskip10pt plus3pt}
\renewcommand{\figurename}{\bf Figure}
\renewcommand{\tablename}{\bf Table}
\newcommand{\mycaption}[1]{\caption{\sl #1}}

\def\mathswitch#1{\relax\ifmmode#1\else$#1$\fi}
\def\mathswitchr#1{\relax\ifmmode{\mathrm{#1}}\else$\mathrm{#1}$\fi}
\newcommand{\PW}{\mathswitchr W}
\newcommand{\PZ}{\mathswitchr Z}
\newcommand{\PH}{\mathswitchr H}

\newcommand{\Pb}{\mathswitchr b}
\newcommand{\Pt}{\mathswitchr t}

\newcommand{\MW}{\mathswitch {M_\PW}}
\newcommand{\GW}{\mathswitch {\Gamma_\PW}}
\newcommand{\MZ}{\mathswitch {M_\PZ}}
\newcommand{\GZ}{\mathswitch {\Gamma_\PZ}}
\newcommand{\MH}{\mathswitch {M_\PH}}

\newcommand{\mb}{\mathswitch {m_\Pb}}
\newcommand{\mt}{\mathswitch {m_\Pt}}

\newcommand{\scrs}{\scriptscriptstyle}
\newcommand{\sw}{\mathswitch {s_{\scrs\PW}}}
\newcommand{\cw}{\mathswitch {c_{\scrs\PW}}}
\newcommand{\mw}{\mathswitch {\overline{M}_\PW}}
\newcommand{\mz}{\mathswitch {\overline{M}_\PZ}}
\newcommand{\gw}{\mathswitch {\overline{\Gamma}_\PW}}
\newcommand{\gz}{\mathswitch {\overline{\Gamma}_\PZ}}
\newcommand{\as}{\alpha_{\mathrm s}}
\newcommand{\at}{\alpha_{\mathrm t}}
\newcommand{\seff}[1]{\sin^2\theta_{\rm eff}^{#1}}

\newcommand{\gev}{\,\, \mathrm{GeV}}

\newcommand{\RR}{{\rm R}}
\newcommand{\LL}{{\rm L}}

\newcommand{\OO}{{\mathcal O}}

\begin{document}

\title{\vspace{-1em}
{\tt \small  \hfill DESY 19-103
\\
\hfill KW 19-005
\\
}
Electroweak pseudo-observables and Z-boson form factors at two-loop accuracy}


\author[a]{Ievgen~Dubovyk,}
\author[b]{Ayres~Freitas,}   \author[c,d]{Janusz~Gluza,}  
\author[c,e]{Tord~Riemann,}  \author[f]{Johann~Usovitsch}
 
\affiliation[a]{II. Institut f{\"u}r Theoretische Physik, Universit{\"a}t Hamburg, 22761 Hamburg,  Germany} 
\affiliation[b]{Pittsburgh Particle Physics, Astrophysics \& Cosmology Center (PITT PACC), Department of Physics \& Astronomy, University of Pittsburgh, Pittsburgh, PA 15260, USA}
\affiliation[c]{Institute  of Physics, University of Silesia, Katowice, Poland }
\affiliation[d]{Faculty of Science, University of Hradec Kr\'alov\'e, Czech Republic}
\affiliation[e]{Deutsches Elektronen--Synchrotron, DESY, 15738 Zeuthen, Germany} 
\affiliation[f]{%
School of Mathematics, Trinity College Dublin, University of Dublin, Dublin 2, Ireland}

\emailAdd{e.a.dubovyk@gmail.com}
\emailAdd{afreitas@pitt.edu}
\emailAdd{gluza@us.edu.pl}
\emailAdd{tordriemann@gmail.com}
\emailAdd{jusovitsch@gmail.com}

\date{today}

\abstract{%
We present Standard Model predictions for the complete set of phenomenologically relevant electroweak precision pseudo-observables related to the $Z$-boson:
the leptonic and bottom-quark
effective weak mixing angles
$\seff{\ell}$, $\seff{b}$, 
the $Z$-boson partial
decay widths $\Gamma_f$, where $f$ indicates any charged lepton, neutrino and quark flavor (except for the top quark), as well as
the total $Z$ decay width $\Gamma_Z$, the 
branching ratios $R_\ell$, $R_c$, $R_b$, and the hadronic cross section $\sigma_{\rm had}^0$.
The input parameters are the masses $\MZ$, $\MH$ and $\mt$, and the couplings $\as$, $\alpha$.
The scheme dependence due to the choice of  $\MW$ or its alternative $G_\mu$
as a last input parameter is also discussed.
Recent substantial technical progress in the calculation of Minkowskian massive higher-order Feynman integrals 
allows the calculation of the complete electroweak two-loop radiative corrections to all the observables mentioned. 
QCD contributions are included appropriately. Results are provided in terms of simple and convenient parameterization formulae whose coefficients have been determined from the full numerical multi-loop calculation.
The size of the missing electroweak three-loop or QCD higher-order corrections is estimated.
We briefly comment on the prospects for their calculation. 
Finally, 
direct predictions for the $Z{\bar f}f$ vector and axial-vector form-factors are given, including a discussion of separate order-by-order contributions.
}

\maketitle
\clearpage

\section{Introduction\label{sec-intro}}

In 2018 we celebrated 50 years of the Standard Model of elementary particles. 
The basics of the model were formulated and experimentally validated in the 1960s/70s. The next decade brought an intensive development of the calculation of quantum field theoretical radiative corrections in that model and in its alternatives.
An experimental highlight in this context was the $e^+e^-$-collider LEP, which enabled us to check the Standard Model at an accuracy of better than the per-cent level, which corresponds to effects from more than one electroweak and two QCD loop orders.
This proved, for the first time in a systematic way, the Standard Model as a quantum field theory.
LEP~1 was running, from Summer 1989 to 1995, at and around the $Z$-boson peak. 
The expectation for the experimental precision of $\MZ$ and $\GZ$ was 20~MeV in 1986 \cite{Blondel:1986kj} and reached finally 2~MeV \cite{Arduini:1996pp}.
This precision tag was extremely important because $\MZ$ is one of the Standard Model input parameters to the commonly used on-mass-shell renormalization scheme. 
Indeed, the experimental accuracy of $\MZ$ triggered much of the precision loop calculations, including the prediction of the top quark and Higgs masses prior to their discoveries from loop corrections to LEP observables in the Standard Model, 
see Refs.~\cite{Ellis:1986jba,Altarelli:1989YR,Bardin:1995-YR03} (as well as Refs.~\cite{Blondel:2018mad,Blondel:2019vdq} for an overview of the current state of the art).
Data from the $Z$ peak and the $Z$ resonance curve (the $Z$ line shape) allow to measure a large variety of observables, such as $\MZ$, $\GZ$, cross-sections for different two-fermion final states and their ratios and angular asymmetries, together with radiation of (sufficiently soft) photons, gluons, etc.
From the real observables,
the so-called electroweak pseudo-observables (EWPOs) are extracted by means of a de-convolution of initial-state radiation and subtraction of backgrounds.
The fine details of relating EWPOs to real cross-sections at LEP~1 precision are described in detail in Ref.~\cite{Gluza:inYR2018} and references quoted therein.

On occasion of the  50$^{th}$ anniversary of the Standard Model, many of the related crucial developments were remembered at the conference "SM@50" \cite{smat50:2019}. 

\bigskip 

This article is devoted to the state-of-the art calculation of the Standard Model (SM) corrections to the $Z{\bar f}f$-vertex and their inclusion into the predictions for the various EWPOs.
We will mostly focus on recent advances in the calculation of the electroweak two-loop terms.
QCD contributions, which are known up to four-loop order, have also been taken into account in the results presented here, but we will refer to the literature for further details. 

The $Z$ resonance curve can be described theoretically by writing the S-matrix elements as a Laurent series in the center-of-mass energy squared $s$ (also called S-matrix ansatz). 
This Laurent series ansatz is worked out up to two loops 
\cite{
Consoli:1986kr,
Sirlin:1991fd,
Willenbrock:1991hu,
Stuart:1991xk,
Sirlin:1991rt,
Stuart:1991cc,
Stuart:1992jf,%
Veltman:1992tm,
Passera:1998uj}.
The coefficients of the leading series term contain the $Z$ vertex form factors.
Their one-loop corrections were studied in the 1980s; first with massless fermions 
\cite{Bardin:1980fe,Bardin:1981sv,Marciano:1980pb,Marciano:1983wwa}, and slightly later with the full mass dependence of the Standard Model
\cite{Akhundov:1985fc,Bernabeu:1987me,Jegerlehner:1988ak,Beenakker:1988pv}.
After several papers on approximate/partial higher-order corrections, 
the complete two-loop weak corrections were determined in a series of papers from 2004 to 2018
\cite{Awramik:2004ge,
Awramik:2006ar,Awramik:2006uz,Hollik:2005va,Hollik:2006ma,Awramik:2008gi,Freitas:2013dpa,Freitas:2014owa,Freitas:2014hra,
Dubovyk:2016aqv,%
Dubovyk:2018rlg}. 
The correct formulation of the interplay of the 2$\to$2 loop corrections with higher order real QED corrections in the S-matrix approach, also called un-folding of the effective 2$\to$2 Born terms from the realistic  2$\to$n observables, is a topic on its own. It was first studied in Refs.~\cite{Leike:1991pq,Riemann:1992gv,Kirsch:1994cf,Riemann:2015wpn}, but its extension beyond two-loop level 
will require more work
\cite{Gluza:inYR2018-C3,Gluza:inYR2018,Jadach:2019bye}.
The numerically relevant two-loop and partial higher-order corrections were included in the final analysis of LEP 1 data \cite{ALEPH:2005ab}.%
\footnote{The EWPOs at LEP~1 were determined order by order without a Laurent expansion.
This was based on the ZFITTER software \cite{Bardin:1999yd,Arbuzov:2005ma}, for both the Standard Model loop calculation and the unfolding of cross-sections. 
The relevant higher-order corrections to the input $W$ mass \cite{Awramik:2003rn} and to the leptonic weak mixing angle \cite{Awramik:2004ge} are 
 implemented in ZFITTER v.6.42. While ZFITTER v.6.44beta  \cite{Akhundov:2013ons} also contains the QCD four-loop corrections of \cite{Baikov:2012er}, they are  of no experimental relevance so far.}

The theoretical advances described here go beyond the Standard Model theory used for physics at LEP~1 \cite{Bardin:1999yd,Arbuzov:2005ma} but will be needed for the FCC-ee Tera-Z project 
\cite{Jadach:April2018,Blondel:2018mad,Mangano:2018mur,Blondel:2019vdq,%
Dubovyk:inYR2018} 
whose unique experimental precision calls for perturbative predictions at three electroweak loops together with corresponding QCD terms.

\bigskip

In this work, the following pseudo-EWPOs are discussed: The partial widths $\Gamma_f$ for $Z$-boson decay into $f\bar{f}$ final states; the total $Z$ width $\GZ$; the branching fractions $\Gamma_f/\GZ$; the total hadronic $Z$-pole quark-pair production cross-section $\sigma^0_{\rm had}$; and the effective weak mixing angles $\seff{f}$, defined from the ratio of vector- and axial $Z$-boson couplings. 
The precise definition of all these pseudo-observables will be given below. 
Pseudo-observables differ from real observables by removing from the former the effects of initial-state and initial-final QED radiation, as well as non-resonant photon-exchange, box and $t$-channel contributions \cite{ALEPH:2005ab,Blondel:2018mad}.

The Standard Model predictions for $Z$-pole pseudo-observables can be constructed in terms of the following three theoretical building blocks \cite{Awramik:2006uz}: 
\begin{align}
v_f(s) &\equiv v_f^\PZ(s) - v_f^\gamma(s)\, 
 \frac{\Sigma_{\rm \gamma Z}(s)}{s+\Sigma_{\gamma\gamma}(s)}\,, \\
a_f(s) &\equiv a_f^\PZ(s) - a_f^\gamma(s)\, 
 \frac{\Sigma_{\rm \gamma Z}(s)}{s+\Sigma_{\gamma\gamma}(s)}\,, \\
\Sigma_\PZ(s) &\equiv \Sigma_{\rm ZZ}(s) - \frac{[\Sigma_{\rm \gamma
Z}(s)]^2}{s+\Sigma_{\gamma\gamma}(s)}\,,
\end{align}
where $v_f^\PZ$ and $a_f^\PZ$ are the one-particle irreducible $Zf\bar{f}$
vector- and axial-vector vertex contributions, respectively, whereas $v_f^\gamma$
and $a_f^\gamma$ are their counterparts for the $\gamma f\bar{f}$ vertex. 
The $\Sigma_{V_1V_2}$ denotes the one-particle irreducible $V_1$--$V_2$ self-energy. At tree level,
\begin{align}
    v^\PZ_{f(0)} &= e\frac{I^3_f - 2Q_f\sw^2}{2\sw\cw}, &
    v^\gamma_{f(0)} &= eQ_f, \\
    a^\PZ_{f(0)} &= e\frac{I^3_f}{2\sw\cw}, &
    a^\gamma_{f(0)} &= 0.
\end{align}
Here $I^3_f$ and $Q_f$ are the weak isospin and electric charge (in units of the elementary charge $e>0$) of the fermion $f$, respectively. $\sw$ and $\cw$ are the sine and cosine of the weak mixing angle, respectively, and the subscript $(0)$ is used to denote tree-level order.

For the theory calculations, these building blocks must be evaluated at the complex $Z$ pole 
\cite{Willenbrock:1991hu,Sirlin:1991fd,Stuart:1991xk,Veltman:1992tm}, $s_0 \equiv \mz^2-i\mz\gz$, where $\mz$ and $\gz$ are the on-shell mass and width of the $Z$-boson, respectively. Note that $\mz$ and $\gz$ differ from the mass $\MZ$ and width $\GZ$ reported in publications of LEP, Tevatron and LHC experiments by a fixed factor \cite{Bardin:1988xt,Leike:1991pq}:
\begin{align}
\textstyle
\mz &= \MZ\big/\sqrt{1+\Gamma_\PZ^2/\MZ^2}\,, \notag \\
\gz &= \Gamma_\PZ\big/\sqrt{1+\Gamma_\PZ^2/\MZ^2}\,. 
\label{massrel}
\end{align}
Similar expressions hold for $\mw$ and $\gw$ \cite{Bardin:1986fi,Denner:1990tx}.

\section{\boldmath $Z$-boson decay width, branching ratios and cross-sections}
\label{sec:zdec}
The width of the $Z$ boson, $\gz$,  is related to the imaginary part of the $Z$ self-energy. Using the optical theorem, one can derive the following expression for $\gz$~
\cite{Freitas:2013dpa,Freitas:2014hra}:
\begin{align}
\gz &= \sum_f \overline{\Gamma}_f, 
~~~ f=e, \mu, \tau, \nu_e, \nu_\mu, \nu_\tau, u,c,s,d,b,
\\ 
\overline{\Gamma}_f &= \frac{N_c^f\mz}{12\pi} \Bigl [
 {\cal R}_{\rm V}^f F_{\rm V}^f + {\cal R}_{\rm A}^f F_{\rm A}^f \Bigr ]_{s=\mz^2} 
 \;. \label{eq:gz}
\end{align}
Here $N_c^f$ is the color factor and ${\cal R}_{\rm V,A}$ are radiator functions that capture final-state QCD and QED corrections, see section 7 in Ref.~\cite{Chetyrkin:1994js}, whereas the remaining electroweak and mixed electroweak--QCD corrections are contained in the form factors $F_{\rm V,A}^f$.
Up to two-loop accuracy, the form factors can be written as follows \cite{Freitas:2014hra}:
\begin{align}
    F_{\rm V}^f = \frac{|v_f|^2}{1+\text{Re}\{\Sigma'_\PZ - \frac{i}{2}\mz\gz \Sigma''_\PZ\}}\bigg|_{s=\mz^2}\,, \\
    F_{\rm A}^f = \frac{|a_f|^2}{1+\text{Re}\{\Sigma'_\PZ - \frac{i}{2}\mz\gz \Sigma''_\PZ\}}\bigg|_{s=\mz^2}\,,
\end{align}
where $\Sigma'_\PZ$ and $\Sigma''_\PZ$ are shorthand expressions for $d\Sigma_\PZ/ds$ and $d^2\Sigma_\PZ/ds^2$, respectively.

In addition to the partial widths, certain branching ratios are of phenomenological importance:
\begin{align}
    R_\ell &= \overline{\Gamma}_{\rm had}/\overline{\Gamma}_\ell, &
    R_c &= \overline{\Gamma}_c/\overline{\Gamma}_{\rm had}, &
    R_b &= \overline{\Gamma}_b/\overline{\Gamma}_{\rm had}.
\end{align}
Here
$\overline{\Gamma}_{\rm had} = 
\sum_{f=u,c,d,s,b} \overline{\Gamma}_{f}$.
Further, the cross-section for $e^+e^- \to \text{hadrons}$ at the $Z$ peak can be expressed in terms of partial widths \cite{Freitas:2014hra},
\begin{align}
   \sigma_{\rm had}&= \sigma^0_{\rm had} + \sigma_{\rm had,non-res}\,, &
    \sigma^0_{\rm had} &= \sum_{f=u,d,c,s,b}
 \frac{12\pi}{\mz^2}\,\frac{\overline{\Gamma}_e\overline{\Gamma}_f}{\gz^2}
 (1+\delta X) \,. \label{eq:xsec}
\end{align}
Here $\sigma^0_{\rm had,non-res}$ accounts for non-resonant photon-exchange, box and $t$-channel contributions. Furthermore, $\delta X$ occurs from higher-order terms of the Laurent expansion of the full amplitude around the complex pole $s_0$. At
two-loop order, $\delta X$ can be written as $\delta X_{(2)} = -(\text{Im}\,\Sigma'_{\PZ(1)})^2 - 2\gz\mz \;\text{Im}\,\Sigma''_{\PZ(1)}$, where subscripts $(n)$ indicate the loop order. In the limit $m_f \ll \MW$ ($f\neq t$), it is given by
\begin{align}
    \delta X_{(2)} &= -\biggl[ \frac{\alpha}{\sw^2\cw^2}\Bigl(\frac{7}{8} - \frac{5}{3}\sw^2+ \frac{20}{9}\sw^4\Bigr)\biggr]^2.
\end{align}
It is important to note that eq.~\eqref{eq:xsec} assumes that initial-state photon radiation effects have been removed by means of a de-convolution procedure, see e.g.\ Ref.~\cite{Schael:2013ita}.

\begin{table}[p]
\renewcommand{\arraystretch}{1.3}
\small
\begin{tabular}{|l|rr@{\hspace{.7em}}r@{\hspace{.7em}}r@{\hspace{.7em}}r@{\hspace{.7em}}rrrr|}
\hline
Observable & \multicolumn{1}{c}{$X_0$} & 
 \multicolumn{1}{c}{$a_1$} & \multicolumn{1}{c}{$a_2$} & 
 \multicolumn{1}{c}{$a_3$} & \multicolumn{1}{c}{$a_4$} & 
 \multicolumn{1}{c}{$a_5$} & \multicolumn{1}{c}{$a_6$} & 
 \multicolumn{1}{c}{$a_7$} & \multicolumn{1}{c|}{$a_8$} \\
\hline
$\Gamma_{e,\mu}$ [MeV] &
 83.983 & $-$0.1202 & $-$0.06919 & 0.00383 & 0.0597 & 0.8037 & $-$0.015 & $-$0.0195 & 0.0032 \\
$\Gamma_{\tau}$ [MeV] &
 83.793 & $-$0.1200 & $-$0.06905 & 0.00382 & 0.0596 & 0.8023 & $-$0.015 & $-$0.0195 & 0.0032 \\
$\Gamma_{\nu}$ [MeV] &
 167.176 & $-$0.1752 & $-$0.1249 & 0.00595 & 0.1046 & 1.253 & $-$0.110 & $-$0.0232 & 0.0064 \\
$\Gamma_{u}$ [MeV] &
 299.994 & $-$0.6152 & $-$0.2771 & 0.0174 & 0.2341 & 4.051 & $-$0.467 & $-$0.0676 & 0.017 \\
$\Gamma_{c}$ [MeV] &
 299.918 & $-$0.6152 & $-$0.2771 & 0.0174 & 0.2340 & 4.051 & $-$0.467 & $-$0.0676 & 0.017 \\
$\Gamma_{d,s}$ [MeV] &
 382.829 & $-$0.6685 & $-$0.3322 & 0.0193 & 0.2792 & 3.792 & $-$0.18 & $-$0.0706 & 0.020 \\
$\Gamma_{b}$ [MeV] &
 375.890 & $-$0.6017 & $-$0.3158 & 0.0190 & 0.227 & $-$2.174 & 0.042 & $-$0.027 & 0.021 \\
$\GZ$ [MeV] & 
 2494.75 & $-$4.055 & $-$2.117 & 0.122 & 1.746 & 19.68 & $-$1.63 & $-$0.432 & 0.12 \\
\hline
$R_\ell$ [$10^{-3}$] &
 20751.6 & $-$8.112 & $-$1.174 & 0.155 & 0.16 & $-$37.59 & $-$10.9 & 1.27 & 0.29 \\
$R_c$ [$10^{-5}$] &
 17222.2 & $-$4.049 & $-$0.749 & 0.0832 & 1.08 & 98.956 & $-$15.1 & $-$0.761 & 0.080 \\
$R_b$ [$10^{-5}$] &
 21585.0 & 4.904 & 0.9149 & $-$0.0535 & $-$2.676 & $-$292.21 & 20.0 & 1.97 & $-$0.11 \\
\hline
$\sigma^0_{\rm had}$ [pb] & 
 41489.6 & 0.408 & $-$0.320 & 0.0424 & 1.32 & 60.17 & 16.3 & $-$2.31 & $-$0.19 \\
\hline
\end{tabular}\\[1ex]
\begin{tabular}{|l|r@{\hspace{.7em}}r@{\hspace{.7em}}rr@{\hspace{.7em}}rrrr|c|}
\hline
Observable & \multicolumn{1}{c}{$a_9$} & \multicolumn{1}{c}{$a_{10}$} & 
 \multicolumn{1}{c}{$a_{11}$} & \multicolumn{1}{c}{$a_{12}$} & 
 \multicolumn{1}{c}{$a_{13}$} & \multicolumn{1}{c}{$a_{14}$} & 
 \multicolumn{1}{c}{$a_{15}$} & \multicolumn{1}{c|}{$a_{16}$} & max.\ dev. \\
\hline
$\Gamma_{e,\mu}$ [MeV] &
 $-$0.0956 & $-$0.0078 & $-$0.0095 & 0.25 & $-$1.08 & 0.056 & $-$0.37 & 286 & $<0.0015$\\
$\Gamma_{\tau}$ [MeV] &
 $-$0.0954 & $-$0.0078 & $-$0.0094 & 0.25 & $-$1.08 & 0.056 & $-$0.37 & 285 & $<0.0015$\\
$\Gamma_{\nu}$ [MeV] &
 $-$0.187 & $-$0.014 & $-$0.014 & 0.37 & $-$0.085 & 0.054 & $-$0.30 & 503 & $<0.002$\\
$\Gamma_{u}$ [MeV] &
 14.26 & 1.6 & $-$0.046 & 1.82 & $-$11.1 & 0.16 & $-$1.0 & 1253 & $<0.006$\\
$\Gamma_{c}$ [MeV] &
 14.26 & 1.6 & $-$0.046 & 1.82 & $-$11.1 & 0.16 & $-$1.0 & 1252 & $<0.006$\\
$\Gamma_{d,s}$ [MeV] &
 10.20 & $-$2.4 & $-$0.052 & 0.71 & $-$10.1 & 0.16 & $-$0.92 & 1469 & $<0.007$\\
$\Gamma_{b}$ [MeV] &
 10.53 & $-$2.4 & $-$0.056 & 1.2 & $-$10.1 & 0.15 & $-$0.95 & 1458 & $<0.007$\\
$\GZ$ [MeV] & 
 58.61 & $-$4.0 & $-$0.32 & 8.1 & $-$56.1 & 1.1 & $-$6.8 & 9267 & $<0.04$\\
\hline
$R_\ell$ [$10^{-3}$] &
 732.30 & $-$44 & $-$0.61 & 5.7 & $-$358 & $-$4.7 & 37 & 11649 & $<0.12$ \\
$R_c$ [$10^{-5}$] &
 230.9 & 125 & 0.045 & 36.9 & $-$120 & 1.2 & $-$6.2 & 3667 & $<0.1$ \\
$R_b$ [$10^{-5}$] &
 $-$131.9 & $-$84 & $-$0.27 & 4.4 & 71.9 & $-$0.77 & $-$4.4 & $-$1790 & $<0.12$ \\
\hline
$\sigma^0_{\rm had}$ [pb] & 
 $-$579.58 & 38 & 0.010 & 7.5 & 85.2 & 9.1 & $-$68 & $\!\!\!-$85957 & $<0.15$\\
\hline
\end{tabular}
\mycaption{Coefficients for the parameterization formula \eqref{par2} for various
observables. 
Within the ranges $25\gev < \MH < 225\gev$, $155\gev < \mt <
195\gev$,
$\as=0.1184\pm 0.0050$, $\Delta\alpha = 0.0590 \pm 0.0005$ and $\MZ = 91.1876 \pm
0.0084 \gev$, the formulae approximate the full result with maximal deviations
given in the last column.
\label{tab:fitr1}}
\end{table}
\afterpage{\clearpage}

Results for the partial and total $Z$ widths, branching ratios and $\sigma^0_{\rm had}$ including the full two-loop corrections have first been published in Ref.~\cite{Dubovyk:2018rlg}.  
They can be expressed in simple parameterization formulae, which are adequate for most phenomenological applications. 
Here, we present slightly more complicated formulae that cover an extended numerical range of input parameters:%
\footnote{This extended input parameter range is useful for determining indirect constraints on various SM parameters from electroweak precision observables (see e.g.\ section 10 in Ref.~\cite{Patrignani:2016xqp}), since these indirect bounds often extend over larger intervals than the corresponding direct measurements.} 
\begin{align}
    &25\gev < \MH < 225\gev, &
    &155\gev < \mt <192\gev,      \notag \\               
    &\MZ = 91.1876 \pm 0.0084\gev, \notag \\
    &\as = 0.1184 \pm 0.0050, &
    &\Delta\alpha = 0.0590 \pm 0.0005.
    \label{eq:parreg}
\end{align}
Here $\Delta\alpha$ is the shift in the running electromagnetic coupling $\alpha(q^2)$ from $q^2=0$ to $\MZ^2$, defined by $\alpha(\MZ^2) = \alpha(0)/(1-\Delta\alpha)$.
It can be divided into a leptonic and a hadronic part, $\Delta\alpha = \Delta\alpha_{\rm lept} + \Delta\alpha_{\rm had}$. $\Delta\alpha_{\rm lept}$ has been computed to three-loop order  \cite{Steinhauser:1998rq}, whereas $\Delta\alpha_{\rm had}$ contains non-perturbative hadronic contributions, which are commonly extracted from data \cite{Davier:2017zfy,Jegerlehner:2017zsb,Keshavarzi:2018mgv}.
We neglect the light fermion masses $m_f$, $f\neq t$, everywhere besides in $\Delta\alpha$ and (at leading power) in the radiator functions ${\cal R}^f_{\rm V/A}$.
The $W$ boson mass
$\MW$ can be computed from the Fermi constant $G_\mu$ (see eqs.~(6)--(8) in the arXiv version of Ref.~\cite{Awramik:2003rn}) and thus is not listed as an independent input parameter. Both $G_\mu$ and $\alpha$, the electromagnetic fine structure constant in the Thomson limit, are known with very small uncertainties, and thus we use their central experimental values \cite{Patrignani:2016xqp} without any uncertainty interval.

The fitting formulae for the EWPOs have the form
\begin{align}
&\begin{aligned}[b] 
X = &X_0 + a_1 L_\PH + a_2 L_\PH^2 + a_3 L_\PH^4 + a_4 \Delta_\PH
 + a_5 \Delta_\Pt + a_6 \Delta_\Pt^2 + a_7 \Delta_\Pt L_\PH + a_8
  \Delta_\Pt L_\PH^2  \\
 & + a_9 \Delta_{\as} + a_{10} \Delta_{\as}^2 + a_{11} \Delta_{\as} \Delta_\PH +
  a_{12} \Delta_{\as} \Delta_\Pt + a_{13} \Delta_\alpha 
  + a_{14} \Delta_\alpha \Delta_\PH \\
  & + a_{15} \Delta_\alpha \Delta_t 
   + a_{16} \Delta_\PZ, 
 \end{aligned} \label{par2} 
 \\[1ex]
&L_\PH = \log\frac{\MH}{125.7\gev}, \quad
 \Delta_\PH = \frac{\MH}{125.7\gev}-1,\quad
 \Delta_\Pt = \Bigl (\frac{\mt}{173.2\gev}\Bigr )^2-1, \quad
 \nonumber \\
& \Delta_{\as} = \frac{\as(\MZ)}{0.1184}-1, \quad
 \Delta_\alpha = \frac{\Delta\alpha}{0.059}-1, \quad
 \Delta_\PZ = \frac{\MZ}{91.1876\gev}-1. \nonumber
\end{align}
The coefficients $X_0$ and $a_1,\ldots a_{16}$ are obtained from fits to a grid of 8750 data points of the full computation. 
The latter includes
\begin{itemize}
    \item Complete one-loop corrections \cite{Akhundov:1985fc}, which have been re-computed for this work, 
    and full two-loop \cite{Freitas:2013dpa,Freitas:2014hra,Dubovyk:2018rlg} electroweak corrections;
    \item Corrections of order $\OO(\alpha\as)$ to vector-boson self-energies \cite{Djouadi:1987gn,Djouadi:1987di,Kniehl:1989yc,Kniehl:1991gu,Djouadi:1993ss}, which have been re-evaluated for this work;
    \item Non-factorizable $\OO(\alpha\as)$ $Zq\bar{q}$ vertex contributions \cite{Czarnecki:1996ei,Harlander:1997zb,Fleischer:1992fq,Buchalla:1992zm,Degrassi:1993ij,Chetyrkin:1993jp}, 
    which are not captured in the products ${\cal R}_i^fF_i^f$ ($i=$ V,A);
    \item Higher-loop QCD corrections in the limit of a large top Yukawa coupling $y_\Pt$, of orders $\OO(\at\as^2)$ \cite{Avdeev:1994db,Chetyrkin:1995ix}, $\OO(\at^2\as)$, $\OO(\at^3)$ 
\cite{vanderBij:2000cg,Faisst:2003px}, and $\OO(\at\as^3)$ \cite{Schroder:2005db,Chetyrkin:2006bj,Boughezal:2006xk}, where $\at \equiv y_\Pt^2/(4\pi)$.
    \item Final-state QED and QCD radiation effects, which enter through the radiator functions ${\cal R}_{\rm V,A}$, up to the orders $\OO(\alpha^2)$, $\OO(\alpha\as)$ and $\OO(\as^4)$ 
\cite{Chetyrkin:1994js,Baikov:2008jh,Kataev:1992dg}.
\end{itemize}
Both the $Z$ vertex corrections as well as the prediction of $\MW$ from $G_\mu$ have been computed to this same level of perturbation theory.

Numerical values for the coefficients are given in Tab.~\ref{tab:fitr1}. Some of the numbers for $X_0$ deviate slightly in the last digit from those in Ref.~\cite{Dubovyk:2018rlg}. This is due to the larger grid of input parameters used here, which can exert a pull on the fit parameters.
The differences are well within the accuracy quoted in the last column of Tab.~\ref{tab:fitr1}.

\section{Asymmetries and effective weak mixing angles}
The effective weak mixing angle for the $Zf\bar{f}$ vertex is defined, from the theory side, as
\begin{align}
    \seff{f}\equiv \frac{1}{4|Q_f|} \biggl(1-\text{Re}\biggl\{\frac{v_f(\mz^2)}{a_f(\mz^2)}\biggr\} \biggr).
\end{align}
Here $\mw^2$ and $\mz^2$ are the real parts of the complex pole of the $W$ and $Z$ propagators, respectively. They are related to the masses commonly reported by experiments at LEP, Tevatron, LHC according to eq~\eqref{massrel}. Moreover,
$Q_f$ denotes the electric charge of the fermion $f$.

The effective weak mixing angles can be extracted from a range of asymmetries \cite{Gluza:inYR2018},
defined from effective Born two-particle cross-sections, 
including the left-right asymmetry
\begin{align}
    A_{\rm LR} &= \frac{\sigma_{e_\LL}-\sigma_{e_\RR}}{\sigma_{e_\LL}+\sigma_{e_\RR}} = A_e + A_{\rm LR}^{\rm non-res}
\intertext{and the forward-backward asymmetry}
    A_{\rm FB}^f &= \frac{\sigma_{\cos\theta>0}-\sigma_{\cos\theta<0}}{\sigma_{\cos\theta>0}+\sigma_{\cos\theta<0}} = \tfrac{3}{4}A_eA_f + A_{\rm FB}^{f,\rm non-res}\,
    ,
\intertext{where}
    A_f &\equiv  \frac{2\,\text{Re}\{v_f/a_f\}}{1+(\text{Re}\{v_f/a_f\})^2}
    = \frac{1-4|Q_f|\seff{f}}{1-4|Q_f|\seff{f}+8(|Q_f|\seff{f})^2}\,.
\end{align}
Here $\sigma_{e_\LL}$ and $\sigma_{e_\RR}$ are the cross-sections for $e^+e^- \to f\bar{f}$ for left- and right-handed polarized electron beams, respectively, whereas $\sigma_{\cos\theta>0}$ and $\sigma_{\cos\theta<0}$ denote the cross-section for $f$ restricted to the forward and backward hemisphere, respectively. Furthermore, $A_X^{\rm non-res}$ accounts for the non-resonant photon-exchange, box and $t$-channel contributions.

The most precisely measured effective weak mixing angles are the leptonic
effective weak mixing angle $\seff{\ell}$ (extracted from $A_{\rm LR}$) and the bottom-quark one, $\seff{b}$ (extracted from $A_{\rm FB}^b$) \cite{Schael:2013ita}.

Standard Model predictions 
for $\seff{\ell}$ including the full two-loop corrections have been presented originally in Ref.~\cite{Awramik:2006ar,Awramik:2006uz,Hollik:2006ma}. 
We reproduced by an independent calculation the contribution of the bosonic electroweak two-loop corrections using the methods of Ref.~\cite{Dubovyk:2018rlg}. The corrections can be expressed in terms of a weak form factor $\Delta\kappa_\ell^{(\alpha^2,\rm bos)}$, where
\begin{align}
\Delta\kappa_f = \biggl(1-\frac{\mw}{\mz}\biggr)^{-1}\seff{f}-1\,.
\end{align}
The comparison with Ref.~\cite{Awramik:2006ar} is shown in Tab.~\ref{tab:compl}, which demonstrates that the two calculations agree to an accuracy of $\OO(10^{-7})$, which implies an accuracy of better than $10^{-7}$ for $\seff{\ell}$.
The full two-loop corrections for $\seff{b}$ have been presented first in Ref.~\cite{Dubovyk:2016aqv}.
\begin{table}[tb]
\centering
\begin{tabular}{|c|c|c|}
\hline
$\MH$ [GeV] & Result of Ref.~\cite{Awramik:2006ar} & Our result \\
\hline
100 & $-0.733989 \times 10^{-4}$ & $-0.733955 \times 10^{-4}$ \\
200 & $-0.469470 \times 10^{-4}$ & $-0.471273 \times 10^{-4}$ \\
\hline 
\end{tabular}
\mycaption{Comparison of numerical results for $\Delta\kappa_\ell^{(\alpha^2,\rm bos)}$ from Ref.~\cite{Awramik:2006ar} with our calculation of $Z$ vertex corrections from Ref.~\cite{Dubovyk:2018rlg}. We use $\mz = 91.1876\gev$ and $\mw=80.385\gev$.
\label{tab:compl}}
\end{table}

In the following, we present simple parameterization formulae for $\seff{\ell}$ and $\seff{b}$, which cover the extended range of input parameters of eq.~\eqref{eq:parreg}.
The parameterization formula 
\begin{align}
&\seff{f} = \begin{aligned}[t] 
 &s_0 + d_1 L_H + d_2 L_H^2 + d_3 L_H^4 + d_4 \Delta_\alpha 
 + d_5 \Delta_\Pt + d_6 \Delta_\Pt^2 + d_7 \Delta_\Pt L_H \label{eq:spar} \\
 &+ d_8 \Delta_{\as} + d_9 \Delta_{\as} \Delta_\Pt + d_{10} \Delta_\PZ 
 \end{aligned}
\intertext{with}
&L_\PH = \log\frac{\MH}{125.7\gev}, \quad
 \Delta_\Pt = \Bigl (\frac{\mt}{173.2\gev}\Bigr )^2-1, \quad
 \nonumber \\
& \Delta_{\as} = \frac{\as(\MZ)}{0.1184}-1, \quad
 \Delta_\alpha = \frac{\Delta\alpha}{0.059}-1, \quad
 \Delta_\PZ = \frac{\MZ}{91.1876\gev}-1 \nonumber
\end{align}
provides a good description of the full result in the parameter region \eqref{eq:parreg}. Values for the coefficients are obtained by fitting \eqref{eq:spar} to a grid of 8750 data points.

Table~\ref{tab:sfit} shows the result of a fit
to a calculation that includes all known corrections:
\begin{itemize}
    \item Complete one- and two-loop electroweak corrections, (see Refs.~\cite{Marciano:1980pb,Akhundov:1985fc,Awramik:2004ge,Hollik:2005va,Awramik:2006ar,Hollik:2006ma,Awramik:2008gi,Dubovyk:2016aqv} for the original references);
    \item Corrections of order $\OO(\alpha\as)$ to vector-boson self-energies \cite{Djouadi:1987gn,Djouadi:1987di,Kniehl:1989yc,Kniehl:1991gu,Djouadi:1993ss}, which we have re-evaluated for this work;
    \item Non-factorizable $\OO(\alpha\as)$ $Zb\bar{b}$ vertex contributions \cite{Czarnecki:1996ei,Harlander:1997zb,Fleischer:1992fq,Buchalla:1992zm,Degrassi:1993ij,Chetyrkin:1993jp}, which do not cancel in the ratio $v_b/a_b$;
    \item Higher-loop corrections in the limit of a large top Yukawa coupling $y_\Pt$, of orders $\OO(\at\as^2)$ \cite{Avdeev:1994db,Chetyrkin:1995ix}, $\OO(\at^2\as)$, $\OO(\at^3)$ 
\cite{vanderBij:2000cg,Faisst:2003px}, and $\OO(\at\as^3)$ \cite{Schroder:2005db,Chetyrkin:2006bj,Boughezal:2006xk} where $\at \equiv y_\Pt^2/(4\pi)$.
\end{itemize}
As indicated by the last column in the table, the largest deviation of the fit formulae from the full result is $\OO(\text{few}\times 10^{-6})$, while for most of the parameter region in \eqref{eq:parreg} the agreement is better than $10^{-6}$.
The careful reader may realize that the parameterization for $\seff{b}$ in Table~\ref{tab:sfit} deviates slightly from Eqs.~(20,22) in \cite{Dubovyk:2016aqv}.
The difference 
is due to the larger grid of data points used here.
A fit formula is, obviously, not able to reproduce the data points in a grid perfectly. 
The fitting aims to find the best average agreement between the data points (which are generated with our full numerical calculation) and the fit formula. A larger grid therefore can lead to some shifts of the coefficients. 
As a consequence, the formula in \cite{Dubovyk:2016aqv} will probably be more accurate for input values within the ranges in Tab.~1 there. On the other hand, while the formula here may be a little less accurate within these ranges, it covers a much larger range of input values. 

\begin{table}[tb]
\centering
\renewcommand{\arraystretch}{1.3}
\setlength{\tabcolsep}{5pt}
\small
\begin{tabular}{|l|rrrrrr|}
\hline
Observable & \multicolumn{1}{c}{$s_0$} & 
 \multicolumn{1}{c}{$d_1$} & \multicolumn{1}{c}{$d_2$} & 
 \multicolumn{1}{c}{$d_3$} & \multicolumn{1}{c}{$d_4$} & 
 \multicolumn{1}{c|}{$d_5$} \\
\hline
$\seff{\ell}\times 10^4$ & 2314.64 & 4.616 & 0.539 & $-$0.0737 & 206 & $-$25.71\phantom{0} \\
$\seff{b}\times 10^4$ & 2327.04 & 4.638 & 0.558 & $-$0.0700 & 207 & $-$9.554 \\
\hline 
\end{tabular} \\[1ex]
\begin{tabular}{|l|rrrrr|c|}
\hline
Observable & \multicolumn{1}{c}{$d_6$} & 
 \multicolumn{1}{c}{$d_7$} & \multicolumn{1}{c}{$d_8$} & 
 \multicolumn{1}{c}{$d_9$} & \multicolumn{1}{c|}{$d_{10}$} &
 max.\ dev. \\
\hline
$\seff{\ell}\times 10^4$ & 4.00 & 0.288 & 3.88 & $-$6.49 & $-$6560 & $<0.056$ \\
$\seff{b}\times 10^4$ & 3.83 & 0.179 & 2.41 & $-$8.24 & $-$6630 & $<0.025$ \\
\hline 
\end{tabular}\hspace{2.2ex}
\mycaption{Coefficients for the parameterization formula \eqref{eq:spar} for the leptonic and bottom-quark effective weak mixing angles. Within the ranges given in eq.~\eqref{eq:parreg}, the formula deviates from the full result up to the maximal amount given in the last column.
\label{tab:sfit}}
\end{table}

It should also be noted that the fit formula for $\seff{\ell}$ in Ref.~\cite{Awramik:2006ar} does not include the $\OO(\at\as^3)$ corrections from Refs.~\cite{Schroder:2005db,Chetyrkin:2006bj,Boughezal:2006xk}, but they are included in the formula presented here.

In Tab.~\ref{tab:futsw} it is shown that the technical accuracy of our fit  formulae is adequate for the expected experimental precision of several future $e^+e^-$ colliders, although it will get modified by anticipated future three-loop electroweak corrections.

\begin{table}[tb]
\centering
\renewcommand{\arraystretch}{1.3}
\setlength{\tabcolsep}{5pt}
\small
\begin{tabular}{|l|ccccc|}
\hline
Observable   & max. dev. & EXP now & FCC-ee & 
  {CEPC} &    
  {GigaZ}   \\
\hline
$\Gamma_Z$ [MeV]   
& 0.04 & 2.3 & 0.1 & 0.5  & 0.8 \\
$\seff{\ell}\times 10^4$  
& 0.056 & 1.6 & 0.06 & 0.23  & 0.1 \\
$\seff{b}\times 10^4$  
& 0.025 &  160 & 9 & 9  & 15 \\
\hline 
\end{tabular} \\[1ex]
\mycaption{Goodness of fit for some chosen EWPOs, compared with the envisaged precision measurements for $\Gamma_Z$ and $\seff{\ell}$ (statistical errors), and $\seff{b}$ (systematic errors) at the collider projects FCC-ee  Tera-Z 
\cite{Abada:2019lih}, CEPC \cite{CEPCStudyGroup:2018ghi} and ILC/GigaZ \cite{Hawkings:1999ac}.
The values of maximal deviations are taken from Tabs.~\ref{tab:fitr1} and \ref{tab:sfit}. The entry ``EXP now'' gives the present experimental precision, as known since LEP~1 \cite{ALEPH:2005ab}.
\label{tab:futsw}}
\end{table}

\section{Vector and axial-vector \boldmath 
$Z$-boson 
form factors $F_{\rm V}^f$ and $F_{\rm A}^f$}
The pseudo-observables discussed in the previous sections aim to be closely related to actual observables, such as cross-sections, branching ratios, or asymmetries. 
On the other hand, for some purposes it is also useful to have numerical results for the underlying vertex corrections themselves \cite{Freitas:2014owa},
for example:
 (i) Inclusion of selected corrections from Beyond Standard Model (BSM) physics, 
 (ii) Estimations of magnitudes of selected single terms, 
(iii) Partial cross-checks with other calculations.
For such purposes, the form factors $F_{\rm V}^f$ and $F_{\rm A}^f$ introduced in eq.~\eqref{eq:gz} are needed explicitly.

Tables~\ref{tab:form} and \ref{tab:formw} show the numerical contributions of different orders of perturbation theory to $F_{\rm V}^f$ and $F_{\rm A}^f$. Here the form factors are always understood to include the appropriate (on-shell) counterterms to render them UV-finite.
The non-factorizable $\OO(\alpha\as)$ corrections are shown separately in the table. Strictly speaking they are not part of the form factors $F_{\rm V}^f$ and $F_{\rm A}^f$, but they finally contribute to observables, such as the decay width, and we incorporate them in the tables and the fit formula below.
In Tab.~\ref{tab:form} the form factors are computed using the following input values:
\begin{subequations}\label{eq:input1}
\begin{align}
    &\MZ = 91.1876\gev, && \GZ = 2.4952\gev, &&\Rightarrow\quad
    \mz = 91.1535\gev \label{eq:input1_z} 
    \\
    &\MW = 80.385\gev, && \GW = 2.085\gev, &&\Rightarrow\quad
    \mw = 80.358\gev \label{eq:input1_w} 
    \\
    &\MH = 125.1\gev, && \mt = 173.2\gev,   \nonumber 
    \\
    & \mb^{\overline{\text{MS}}} = 4.2\gev, 
    && \Delta\alpha = 0.059, && \as = 0.1184
\end{align}
\end{subequations}
For Tab.~\ref{tab:formw}, on the other hand, the Fermi constant $G_\mu$ is used as an input instead of \eqref{eq:input1_w}, and $\mw$ is computed from
\begin{align}
    \mw^2\biggl(1-\frac{\mw^2}{\mz^2}\biggr) = \frac{\pi\alpha}{\sqrt{2}G_\mu}(1+\Delta r),
\end{align}
where $\Delta r$ has been evaluated to the same orders as given in each column of the table. More details about the calculation of $\Delta r$ can be found in Ref.~\cite{Awramik:2003rn}.
As before, the dependence of the Standard Model prediction on various input parameters can be expressed in terms of the simple parameterization formula eq.~\eqref{par2}. 

\begin{table}[t]
\renewcommand{\arraystretch}{1.3}
\footnotesize
\begin{tabular}{|l|rrcccccc|}
\hline &&&&&&&& \\[-1.3em]
Form fact. & \multicolumn{1}{c}{Born} &
 \multicolumn{1}{c}{${\cal O}(\alpha)$} & 
 \multicolumn{1}{c}{${\cal O}(\alpha\as)$} & 
 \parbox{4em}{\centering ${\cal O}(\alpha\as)$\\non-fact.} & 
 \parbox{6.5em}{${\cal O}(\at\as^2,\,\at\as^3,$\\
 ${}\quad\,\at^2\as,\,\at^3)$} & 
 \multicolumn{1}{c}{${\cal O}(N_f^2\alpha^2)$} & 
 \multicolumn{1}{c}{${\cal O}(N_f\alpha^2)$} & 
 \multicolumn{1}{c|}{${\cal O}(\alpha^2_{\rm bos})$} \\[2ex]
\hline
$F_V^\ell$ [$10^{-5}$] &
 39.07 & $-$24.86 & 2.41 & -- & 0.35 & 1.47 & 2.37 & 0.27 \\
$F_A^\ell$ [$10^{-5}$] &
 3309.44 & 118.59 & 9.46 & -- & 1.22 & 8.60 & 2.60 & 0.45 \\
$F_{V,A}^\nu$ [$10^{-5}$] &
 3309.44 & 127.56 & 9.46 & -- & 1.22 & 8.60 & 3.83 & 0.39 \\
$F_V^{u,c}$ [$10^{-5}$] &
 544.88 & $-$44.80 & 7.29 & $-$0.39 & 1.02 & $-$1.67 & 3.25 & 0.33 \\
$F_A^{u,c}$ [$10^{-5}$] &
 3309.44 & 120.79 & 9.46 & $-$0.98 & 1.22 & 8.60 & 3.27 & 0.44 \\
$F_V^{d,s}$ [$10^{-5}$] &
  1635.01 & 5.84 & 9.64 & $-$0.80 & 1.32 & 0.71 & 3.45 & 0.37 \\
$F_A^{d,s}$ [$10^{-5}$] &
  3309.44 & 123.78 & 9.46 & $-$1.14 & 1.22 & 8.60 & 3.11 & 0.42 \\
$F_V^b$ [$10^{-5}$] &
  1635.01 & $-$26.16 & 9.64 & \phantom{$-$}3.13 & 1.32 & 0.71 & 1.77 & 1.05 \\
$F_A^b$ [$10^{-5}$] &
  3309.44 & 78.26 & 9.46 & \phantom{$-$}4.45 & 1.22 & 8.60 & 0.13 & 1.18 \\
\hline
\end{tabular}
\mycaption{Contributions of different perturbative orders to the $Z$ vertex form factors. A 
fixed value of $\MW$ has been used as input, instead of $G_\mu$. $N_f^n$ refers to corrections with $n$ closed fermions loops, whereas $\alpha_{\rm bos}^2$ denotes corrections without closed fermions loops. Furthermore, $\at = y_\Pt/(4\pi)$ where $y_\Pt$ is the top Yukawa coupling.
\label{tab:form}}
\end{table}

\begin{table}[tbp]
\renewcommand{\arraystretch}{1.3}
\footnotesize
\begin{tabular}{|l|rrrcccc|}
\hline &&&&&&& \\[-1.3em]
Form fact. & \multicolumn{1}{c}{Born} &
 \multicolumn{1}{c}{${\cal O}(\alpha)$} & 
 \multicolumn{1}{c}{${\cal O}(\alpha\as)$} & 
 \parbox{4em}{\centering ${\cal O}(\alpha\as)$\\non-fact.} & 
 \parbox{6.5em}{${\cal O}(\at\as^2,\,\at\as^3,$\\
 ${}\quad\,\at^2\as,\,\at^3)$} & 
 \multicolumn{1}{c}{${\cal O}(N_f^2\alpha^2,\,N_f\alpha^2)$} & 
 \multicolumn{1}{c|}{${\cal O}(\alpha^2_{\rm bos})$} \\[2ex]
\hline
$F_V^\ell$ [$10^{-5}$] &
   77.63 &  $-$59.86 &    0.31 & -- & $-$0.09 & 1.88 & 0.24 \\
$F_A^\ell$ [$10^{-5}$] &
 3426.43 &     19.32 & $-$1.12 & -- & $-$0.92 & 1.62 & 0.27 \\
$F_{V,A}^\nu$ [$10^{-5}$] &
 3426.43 &     28.36 & $-$1.16 & -- & $-$0.93 & 2.81 & 0.22 \\
$F_V^{u,c}$ [$10^{-5}$] &
  644.45 & $-$129.87 & $-$1.36 & $-$0.40 & $-$0.73 & $-$6.25 & 0.19 \\
$F_A^{u,c}$ [$10^{-5}$] &
 3426.43 &     21.54 & $-$1.13 & $-$0.99 & $-$0.92 & 2.28 & 0.27 \\
$F_V^{d,s}$ [$10^{-5}$] &
 1760.71 & $-$100.64 & $-$1.85 & $-$0.81 & $-$1.01 & $-$6.23 & 0.19 \\
$F_A^{d,s}$ [$10^{-5}$] &
 3426.43 &     24.56 & $-$1.15 & $-$1.15 & $-$0.93 & 2.11 & 0.25 \\
$F_V^b$ [$10^{-5}$] &
 1760.71 & $-$133.08 & $-$1.58 & \phantom{$-$}3.15 & $-$0.96 & $-$7.70 & 0.86 \\
$F_A^b$ [$10^{-5}$] &
 3426.43 &  $-$21.45 & $-$0.85 & \phantom{$-$}4.47 & $-$0.88 & $-$0.64 & 1.01 \\
\hline
\end{tabular}
\mycaption{Same as Tab.~\ref{tab:form}, but with $\MW$ calculated from $G_\mu$.
\label{tab:formw}}
\end{table}

Table~\ref{tab:fitrf} shows the numerical values for the coefficients obtained 
by fitting this formula to the currently most precise computation, including
the same corrections as in section~\ref{sec:zdec}, except for the final-state QED and QCD radiation effects, i.e.
\begin{align}
    F_{\rm V}^f &= |v_{f(0)}|^2 + F^f_{\rm V(\alpha)} +
    F^f_{\rm V(\alpha\as)} + F^f_{\rm V(\alpha^2)} +
    F^f_{\rm V(\alpha_\Pt\as^2)} +
    F^f_{\rm V(\alpha_\Pt^2\as)} +
    F^f_{\rm V(\alpha_\Pt^3)} +
    F^f_{\rm V(\alpha_\Pt\as^3)}
    \,
    , \notag
    \\
    %
    F_{\rm A}^f &= |a_{f(0)}|^2 + F^f_{\rm A(\alpha)} +
    F^f_{\rm A(\alpha\as)} + F^f_{\rm A(\alpha^2)} +
    F^f_{\rm A(\alpha_\Pt\as^2)} +
    F^f_{\rm A(\alpha_\Pt^2\as)} +
    F^f_{\rm A(\alpha_\Pt^3)} +
    F^f_{\rm A(\alpha_\Pt\as^3)}\,.
\end{align}
Note that $G_\mu$ (rather than $\MW$) has been used as one input in Tab.~\ref{tab:fitrf}.

The form factor results presented here can be easily augmented to include the effects of some new physics model:
\begin{align}
    F^f_{\rm V,SM{+}NP} &\approx F^f_{\rm V,SM} + 2 \text{Re}\{v_{f(0)} v_{f,\rm NP}\}\,, \\
    F^f_{\rm A,SM{+}NP} &\approx F^f_{\rm A,SM} + 2 \text{Re}\{a_{f(0)} a_{f,\rm NP}\}\,.
\end{align}
Here ``SM'' denotes the SM contributions discussed in the present paper, while ``NP'' stands for the new physics correction on top of the SM. Since the existing experimental constraints imply that any possible new physics effect is small, it is sufficient to use the tree-level couplings $v_{f(0)}$ and $a_{f(0)}$ in the interference terms and neglect the $|v_{f,\rm NP}|^2$ and $|a_{f,\rm NP}|^2$ terms.


\begin{table}[tbp]
\renewcommand{\arraystretch}{1.3}
\small
\begin{tabular}{|l|rrrrrrrrr|}
\hline
Form fact. & \multicolumn{1}{c}{$X_0$} & 
 \multicolumn{1}{c}{$a_1$} & \multicolumn{1}{c}{$a_2$} & 
 \multicolumn{1}{c}{$a_3$} & \multicolumn{1}{c}{$a_4$} & 
 \multicolumn{1}{c}{$a_5$} & \multicolumn{1}{c}{$a_6$} & 
 \multicolumn{1}{c}{$a_7$} & \multicolumn{1}{c|}{$a_8$} \\
\hline
$F_V^\ell$ [$10^{-5}$] &
 20.11 & $-$1.317 & $-$0.2614 & 0.0333 & 0.276 & 7.472 & 1.55 & $-$0.326 & 0.0012 \\
$F_A^\ell$ [$10^{-5}$] &
 3445.60 & $-$3.640 & $-$2.592 & 0.124 & 2.186 & 25.67 & $-$2.17 & $-$0.480 & 0.13 \\
$F_{V,A}^\nu$ [$10^{-5}$] &
 3455.73 & $-$3.618 & $-$2.580 & 0.123 & 2.161 & 25.89 & $-$2.27 & $-$0.480 & 0.13 \\
$F_V^{u,c}$ [$10^{-5}$] &
 506.025 & $-$4.483 & $-$1.062 & 0.105 & 0.912 & 23.20 & $-$1.54 & $-$0.409 & 0.085 \\
$F_A^{u,c}$ [$10^{-5}$] &
 3447.47 & $-$3.630 & $-$2.588 & 0.124 & 2.178 & 25.80 & $-$2.24 & $-$0.480 & 0.13 \\
$F_V^{d,s}$ [$10^{-5}$] &
 1650.36 & $-$5.243 & $-$1.834 & 0.134 & 1.540 & 29.20 & $-$2.78 & $-$0.458 & 0.13 \\
$F_A^{d,s}$ [$10^{-5}$] &
 3450.12 & $-$3.612 & $-$2.581 & 0.123 & 2.167 & 25.64 & $-$2.21 & $-$0.476 & 0.13 \\
$F_V^b$ [$10^{-5}$] &
 1621.40 & $-$4.888 & $-$1.756 & 0.134 & 1.27 & $-$2.87 & $-$1.5 & $-$0.21 & 0.13 \\
$F_A^b$ [$10^{-5}$] &
 3408.08 & $-$3.110 & $-$2.458 & 0.120 & 1.754 & $-$21.04 & $-$0.60 & $-$0.15 & 0.14 \\
\hline
\end{tabular}\\[1ex]
\begin{tabular}{|l|rrrrrrrr|c|}
\hline
Form fact. & \multicolumn{1}{c}{$a_9$} & \multicolumn{1}{c}{$a_{10}$} & 
 \multicolumn{1}{c}{$a_{11}$} & \multicolumn{1}{c}{$a_{12}$} & 
 \multicolumn{1}{c}{$a_{13}$} & \multicolumn{1}{c}{$a_{14}$} & 
 \multicolumn{1}{c}{$a_{15}$} & \multicolumn{1}{c|}{$a_{16}$} & max.\ dev. \\
\hline
$F_V^\ell$ [$10^{-5}$] &
 $-$0.13 & $-$0.037 & $-$0.11 & 2.78 & $-$43.15 & 1.2 & $-$8.9 & 1396 & $<0.02$ \\
$F_A^\ell$ [$10^{-5}$] &
 $-$3.87 & $-$0.28 & $-$0.28 & 7.6 & $-$1.9 & 1.1 & $-$6.3 & 6912 & $<0.04$ \\
$F_{V,A}^\nu$ [$10^{-5}$] &
 $-$3.93 & $-$0.29 & $-$0.28 & 7.6 & $-$1.7 & 1.1 & $-$6.3 & 6939 & $<0.04$ \\
$F_V^{u,c}$ [$10^{-5}$] &
 $-$3.89 & $-$0.16 & $-$0.22 & 6.5 & $-$145.6 & 0.98 & $-$6.9 & 5648 & $<0.03$\\
$F_A^{u,c}$ [$10^{-5}$] &
 $-$4.86 & $-$0.27 & $-$0.27 & 7.5 & $-$1.9 & 1.1 & $-$6.3 & 6918 & $<0.04$\\
$F_V^{d,s}$ [$10^{-5}$] &
 $-$5.56 & $-$0.24 & $-$0.28 & 8.1 & $-$132 & 0.95 & $-$6.0 & 7492 & $<0.04$\\
$F_A^{d,s}$ [$10^{-5}$] &
 $-$5.03 & $-$0.27 & $-$0.27 & 7.5 & $-$1.6 & 1.1 & $-$6.2 & 6923 & $<0.04$\\
$F_V^b$ [$10^{-5}$] &
 $-$1.3 & $-$0.27 & $-$0.32 & 12.5 & $-$131 & 0.95 & $-$5.7 & 7451 & $<0.04$\\
$F_A^b$ [$10^{-5}$] &
 0.87 & $-$0.31 & $-$0.31 & 13.5 & $-$2.4 & 1.1 & $-$7.2 & 6918 & $<0.04$\\
\hline
\end{tabular}
\mycaption{Coefficients for the parameterization formula \eqref{par2} for various
form factors ($X$). Within the ranges $25\gev < \MH < 225\gev$, $155\gev < \mt <
195\gev$,
$\as=0.1184\pm 0.0050$, $\Delta\alpha = 0.0590 \pm 0.0005$ and $\MZ = 91.1876 \pm
0.0084 \gev$, the formula approximates the full result with average and maximal deviations
given in the last two columns.
\label{tab:fitrf}}
\end{table}

\afterpage{\clearpage}

\section{Theoretical error estimates for missing higher order corrections \label{sec:therror}}
The main theory uncertainty of the results presented in this paper stems from unknown three- and four-loop corrections. The leading missing orders are $\OO(\alpha^3)$, $\OO(\alpha^2\as)$, $\OO(\alpha\as^2)$, and $\OO(\alpha\as^3)$. Partial results for these contributions, in the limit of a large top Yukawa coupling $y_\Pt$, have already been computed \cite{Avdeev:1994db,Chetyrkin:1995ix,vanderBij:2000cg,Faisst:2003px,Schroder:2005db,Chetyrkin:2006bj,Boughezal:2006xk}. Therefore, when evaluating the impact of theory uncertainties, it is always implied that we refer to these contributions \emph{beyond} the leading-$y_\Pt$ limit.

There are a number of different methods for assessing theory uncertainties from unknown higher orders, none of which is
fully reliable. Rather, they should be taken as an order-of-magnitude estimate of the size of these terms. A convenient and widely applicable method is based on the assumption that the first few orders of the perturbation series approximately follow a geometric series \cite{Freitas:2014hra,Freitas:2016sty,Dubovyk:2018rlg}.
In this way one obtains as an ansatz
\begin{equation}
\begin{aligned} 
\OO(\alpha^3)-\OO(\alpha_\Pt^3) &\sim 
 \frac{\OO(\alpha^2)-\OO(\alpha_\Pt^2)}{\OO(\alpha)}
 \OO(\alpha^2), \\
\OO(\alpha^2\as)-\OO(\alpha_\Pt^2\as) &\sim 
 \frac{\OO(\alpha^2)-\OO(\alpha_\Pt^2)}{\OO(\alpha)}
 \OO(\alpha\as), \\
\OO(\alpha\as^2)-\OO(\alpha_\Pt\as^2) &\sim 
 \frac{\OO(\alpha\as)-\OO(\alpha_\Pt\as)}{\OO(\alpha)}
 \OO(\alpha\as), \\
\OO(\alpha\as^3)-\OO(\alpha_\Pt\as^3) &\sim 
 \frac{\OO(\alpha\as)-\OO(\alpha_\Pt\as)}{\OO(\alpha)}
 \OO(\alpha\as^2),
\end{aligned} \label{errprop}
\end{equation}
where $\at = y_\Pt^2/(4\pi)$. Since we are only interested in the missing higher orders beyond the leading large-$y_\Pt$ limit, the same leading large-$y_\Pt$ approximations have been subtracted in the
numerators on the right-hand sides.

The contribution of these estimates to the overall theory error evaluation is shown in Tab.~\ref{tab:err1} for various pseudo-observables, and in Tab.~\ref{tab:err2} for the $Z$-boson form factors.
Note that the error estimate for $\seff{\ell}$ is slightly improved compared to Refs.~\cite{Awramik:2006ar,Awramik:2006uz} due to the inclusion of $\OO(\at\as^3)$ corrections from Refs.~\cite{Schroder:2005db,Chetyrkin:2006bj,Boughezal:2006xk}.

Nevertheless, we would also like to remind the reader that any estimate of the theory error from missing higher orders is not a precise prediction. Therefore it is generally desirable to ensure that the theory error is sub-dominant in any phenomenological analysis. Comparing the numbers in Tab.~\ref{tab:err1} to current measurement results \cite{Schael:2013ita,Patrignani:2016xqp}, this is clearly seen to be the case.

\begin{table}[tbhp]
\centering
\renewcommand{\arraystretch}{1.3}
\begin{tabular}{|l|cccc|c|}
\hline
Observable & $\alpha\as^2$ & $\alpha\as^3$ & $\alpha^2\as$ & $\alpha^3$ & Total \\
\hline
$\Gamma_{e,\mu,\tau}$ [MeV] & 0.008 & 0.001 & 0.010 & 0.013 & 0.018 \\
$\Gamma_{\nu}$ [MeV] & 0.008 & 0.001 & 0.008 & 0.011 & 0.016 \\
$\Gamma_{u,c}$ [MeV] & 0.025 & 0.004 & 0.08 & 0.07 & 0.11 \\
$\Gamma_{d,s}$ [MeV] & 0.016 & 0.003 & 0.06 & 0.05 & 0.08 \\
$\Gamma_{b}$ [MeV] & 0.11 & 0.02 & 0.13 & 0.06 & 0.18 \\
$\GZ$ [MeV] & 0.23 & 0.035 & 0.21 & 0.20 & 0.4 \\
\hline
$R_\ell$ [$10^{-3}$] & 2.5 & 0.4 & 3.6 & 3.9 & 6 \\
$R_c$ [$10^{-5}$] & 1.6 & 0.3 & 3.4 & 3.0 & 5 \\
$R_b$ [$10^{-5}$] & 5.5 & 0.9 & 6.4 & 3.7 & 10 \\
\hline
$\sigma^0_{\rm had}$ [pb] & 0.2 & 0.03 & 4.2 & 3.7 & 6 \\
\hline
$\seff{\ell}$ [$10^{-5}$] & --- & 0.3 & 3.0 & 3.1 & 4.3 \\
$\seff{b}$ [$10^{-5}$] & 0.7 & 0.4 & 4.3 & 3.2 & 5.3 \\
\hline
\end{tabular}
\mycaption{Leading unknown higher-order corrections and their estimated order of magnitude for various pseudo-observables. The different orders always correspond to missing higher orders beyond the known approximations in the limit of a large top Yukawa coupling. The total theory error is obtained by adding the individual orders in quadrature.
\label{tab:err1}}
\end{table}

\begin{table}[tbhp]
\centering
\renewcommand{\arraystretch}{1.3}
\begin{tabular}{|l|cccc|c|}
\hline
Observable & $\alpha\as^2$ & $\alpha\as^3$ & $\alpha^2\as$ & $\alpha^3$ & Total \\
\hline
$F_V^\ell$ [$10^{-5}$] & 0.03 & 0.004 & 0.2 & 0.4 & 0.5 \\
$F_A^\ell$ [$10^{-5}$] & 0.17 & 0.03 & 0.3 & 0.3 & 0.4 \\
$F_{V,A}^\nu$ [$10^{-5}$] & 0.16 & 0.02 & 0.3 & 0.5 & 0.6 \\
$F_V^{u,c}$ [$10^{-5}$] & 0.09 & 0.01 & 0.4 & 0.2 & 0.5 \\
$F_A^{u,c}$ [$10^{-5}$] & 0.17 & 0.03 & 0.3 & 0.4 & 0.5 \\
$F_V^{d,s}$ [$10^{-5}$] & 0.2 & 0.03 & 0.6 & 0.8 & 1.1 \\
$F_A^{d,s}$ [$10^{-3}$] & 0.3 & 0.04 & 0.4 & 0.5 & 0.7 \\
$F_V^b$ [$10^{-5}$] & 0.2 & 0.03 & 0.8 & 0.7 & 1.1 \\
$F_A^b$ [$10^{-5}$] & 0.3 & 0.04 & 0.1 & 0.1 & 0.3 \\
\hline
\end{tabular}
\mycaption{Same as Tab.~\ref{tab:err1}, but for various form factors.
\label{tab:err2}}
\end{table}                              

\afterpage{\clearpage}

\section{Conclusions and outlook}
In this study, we have presented some phenomenologically useful applications of the recently completed electroweak two-loop calculation of $Z$-boson vertex corrections 
\cite{Dubovyk:2016aqv,Dubovyk:2018rlg}.
The work collects multi-year efforts of several groups for predictions of the
EWPOs related to the $Z$ peak up to electroweak full two-loop accuracy, supplemented by leading QCD higher-order terms.
We have determined the two-loop electroweak contributions with a net relative numerical accuracy of about four digits. 
This ensures that these two-loop results will be
known with sufficient accuracy
even when adding the next perturbative order, 
as it might be needed for applications at the next generation of $e^+e^-$ colliders.

For practical applications, the results for the EWPOs, as well as for the $Z$-boson vertex form factors, have been presented in terms of simple parameterization formulae, whose coefficients have been fitted to the full numerical computation.
It is planned to include these fitting formulae into a new version of the weak library DIZET of ZFITTER \cite{AkhundovDIZET:2019??,Bardin:1989di,Bardin:1989tq,Bardin:1999yd,%
Arbuzov:2005ma,Akhundov:2013ons}.%
\footnote{Private communication by L. Kalinovskaya for the ZFITTER/DIZET support team.}
The accuracy of the fitting formulae is less
than our full numerical two-loop calculation, but more than sufficient for present-day purposes.
For the future FCC-ee Tera-Z project, it may be necessary to provide more precise formulae
by including  more terms with higher powers of the input parameters.

Finally, we would like to make a few comments on the prospects for the calculation of electroweak three-loop corrections, which will be necessary for the level of precision foreseen for FCC-ee and similar $e^+e^-$ collider proposals.
This electroweak third order, by itself, will be needed with
only about two digits accuracy \cite{Dubovyk:inYR2018B1}.
The generation of the amplitudes for ${\cal O}(10^4{-}10^5)$ diagrams as well as the evaluation of the Lorentz and Dirac algebra are routine tasks performed by computer algebra programs, and they should be straightforward with increased computing power in the future.
Potential specific problems related to the treatment of $\gamma_5$ at three-loop level have to be controlled \cite{Marquard:inYR2018D1}.

The most challenging problem will certainly be the stable numerical computation of three-loop Feynman integrals with several different internal mass scales.
At two loops, we did not perform any reduction of the Feynman integrals to a smaller number of masters and thus had to calculate about 1000 previously unknown numerical integrals.
In the next perturbative order, it may be advantageous to perform such a reduction to masters,
given the ever increasing performance of programs like KIRA \cite{Maierhoefer:2017hyi,Maierhofer:2019goc}, Reduze~2 \cite{vonManteuffel:2012np}, FIRE \cite{Smirnov:2014hma,Smirnov:2014hma}, and LiteRED \cite{Lee:2013mka}.

For the calculation of the (master) integrals themselves, it is desirable to have a procedure to automatically isolate and treat the ultra-violet and infra-red singularities.
Although there is rapid progress in several analytical approaches to complicated loop integrals \cite{Blondel:2018mad,Blondel:2019vdq}, one has to expect that the
{more complicated ones will have to be done numerically.
An additional complication is that 
for integrals with physical cuts, their stable numerical evaluation becomes more challenging.}
There are two kinds of software packages available that address these problems,
based on either sector decomposition (SD)
as realized in the SecDec project
\cite{
Carter:2010hi,Borowka:2012yc,Borowka:2015mxa,Borowka:2017idc,Borowka:2018goh,Jahn:2018zsh%
} 
and the FIESTA project \cite{
Smirnov:2008py,Smirnov:2009pb,Smirnov:2013eza,Smirnov:2015mct%
}, 
or based on Mellin-Barnes (MB) transformations, as implemented in the AMBRE project
\cite{
Gluza:2007rt,Gluza:2009mj,Gluza:2010rn,Gluza:2010mz,Blumlein:2014maa,Dubovyk:2015yba,%
Dubovyk:2016ocz,Dubovyk:2017cqw,Dubovyk:inYR2018E4,Dubovyk:PhD2019??,
Usovitsch:2018shx,Usovitsch:April2018,Usovitsch:inYR2018,%
Usovitsch:PhD2018%
}.
Sector decomposition is typically advantageous for integrals with many different mass scales, while the MB approach is more efficient for integrals with fewer independent parameters.
Both methods certainly have room for crucial improvements. 
Several other numerical integration methods, as reviewed e.g.\ in Refs.~\cite{Blondel:2018mad,Blondel:2019vdq}, are useful for certain classes of multi-loop integrals, even though they are less general than the SD and MB approaches. Overall, numerical loop integration techniques are well positioned to meet the necessary future precision demands.

An electroweak three-loop result for the $Z$-peak EWPOs
must be accompanied by improved calculations of the corrections needed to translate \mbox{EWPOs} to real observables. These include initial-state and final-state QED corrections and their interference as well as higher-order terms of the Laurent series expansion about the $Z$ resonance pole \cite{Blondel:2018mad}. The latter will, for example, involve
massive two-loop box diagrams. A complete accounting of the required correction terms is still lacking.

\bigskip 

To summarize,
we have completed the electroweak two-loop predictions for the EWPOs of the $Z$ resonance and collect here an extensive 
set of new fitting formulae for them.
On a longer time scale, the calculation of the next perturbative order for the calculation of the EWPOs will be 
necessary and, with  proper investments, 
realistically accessible.

\acknowledgments   
The authors are grateful to V.~Yermolchyk for checks and useful comments.
The work of \textit{I.D.} was supported by a research grant of Deutscher Akademischer Austauschdienst (DAAD) and by Deutsches Elektronensychrotron DESY and CERN.
The work of \textit{J.G.} was supported by the Polish National Science Centre (NCN) under the Grant Agreement 2017/25/B/ST2/01987 and by 
the program International mobilities for research activities of the University of Hradec Kr\'alov\'e,  CZ.02.2.69/0.0/0.0/\linebreak[0]16\_027/\linebreak[0]0008487.
The work of \textit{A.F.}\  was supported in part by the National Science Foundation under
grant no.\ PHY-1820760.
The work of \textit{T.R.}\ is funded by Deutsche Rentenversicherung Bund. He is supported
in part by a 2015 Alexander von Humboldt Honorary Research Scholarship of the Foundation
for Polish Sciences (FNP) and by the Polish National Science Centre (NCN) under the Grant
Agreement 2017/25/B/ST2/01987. 
 \textit{J.U.} received funding from the European Research Council (ERC) under the European Union’s Horizon
2020 research and innovation programme under grant agreement no. 647356 (CutLoops).
The project was partly supported by the COST (European Cooperation in Science and Technology) Action CA16201 PARTICLEFACE.
We finally acknowledge kind support for a working meeting by Prof. Peter Uwer at Humboldt-Universit{\"a}t zu Berlin.
 
\addcontentsline{toc}{section}{References}

\bibliographystyle{elsarticle-num}

\end{document}